\documentclass{aipproc}

\newcommand{\be}{\begin{equation}}
\newcommand{\ee}{\end{equation}}
\newcommand{\bea}{\begin{eqnarray}}
\newcommand{\eea}{\end{eqnarray}} 

\newcommand{\de}{\partial}

\layoutstyle{6x9}

\SetInternalRegister\hbadness{8000} 

\newcommand\doingARLO[2][]{%
  \ifx\mmref\undefined #1\else #2\fi
}

\begin{document}
\def\esp #1{e^{\displaystyle{#1}}}
\def\slash#1{\setbox0=\hbox{$#1$}#1\hskip-\wd0\dimen0=5pt\advance
       \dimen0 by-\ht0\advance\dimen0 by\dp0\lower0.5\dimen0\hbox
         to\wd0{\hss\sl/\/\hss}}\def\ink {\int~{d^4k\over (2\pi)^4}~}

\title
      [Color superconductivity in high density QCD]
      {Color superconductivity in high density QCD}


\author{Roberto Casalbuoni}{
  address={Dipartimento di Fisica dell' Universita' di Firenze and Sezione
INFN, Via G. Sansone 1, 50019 Sesto Fiorentino (Firenze), Italy.
E-mail: casalbuoni@fi.infn.it},
  email={casalbuoni@fi.infn.it},
}



\date{\today}

\maketitle

It is now a well established fact that at zero temperature and
sufficiently high densities quark matter is a color superconductor
\cite{barrois,cs}. The study starting from first principles was done
in \cite{weak,PR-sp1,weak-cfl}. At densities much higher that the
masses of the quarks $u$, $d$ and $s$, the favored state is the
so-called Color-Flavor-Locking (CFL) state, whereas at lower
densities the strange quark decouples and the relevant phase is
called two-flavor color superconducting (2SC).

An interesting possibility is that in the interior of compact
stellar objects (CSO) some color superconducting phase may exist.
In fact we recall  that the central densities for these stars
could be up to $1\div 1.5$ fm$^{-3}$, whereas the temperature is
of the order of tens of keV. However the usual assumptions leading
to show, for instance, that with three flavors the favored state
is CFL, should now be reviewed. Matter inside a CSO should be
electrically neutral and should not carry any color. Also
conditions for $\beta$-equilibrium should be fulfilled. As far as
color is concerned, it is possible to impose a simpler condition,
that is color neutrality, since in \cite{Amore:2001uf} it has been
shown that there is no free energy cost in projecting color
singlet states out of color neutral states. With the further
observation that using $\beta$-equilibrium one has $\mu_e=-\mu_Q$,
where $\mu_e$ and $\mu_Q$ are the chemical potentials associated
to the electrons and to the electric charge respectively, we see
that all the previous conditions can be satisfied by requiring
that the electric-charge, $T_3$ and $T_8$ densities vanish. This
is equivalent to require that the derivatives of the free energy
with respect to the corresponding chemical potentials are zero:
\be \frac{\de \Omega}{\de \mu_e}=\frac{\de \Omega}{\de \mu_3}=
\frac{\de \Omega}{\de \mu_8}=0\label{eq:1}\ee A color
superconducting state is also characterized by a non-zero
expectation value of a diquark operator which depends, in the
homogeneous case, on several constants, the gaps. The free energy
is evaluated by starting from a microscopic description of the
quark interaction, which is usually assumed to be a four-fermi
interaction. Of course this is an approximation to the real case,
but it turns out to be quite effective. In this way one gets the
free-energy as a function of the chemical potentials and of the
energy gaps, $\Delta_i$. Therefore, besides requiring the
conditions (\ref{eq:1}), we have  to minimize the free energy also
with respect to $\Delta_i$ \be \frac{\de \Omega}{\de
\Delta_i}=0\ee Another important point to be considered in these
applications is the mass of the strange quark, since the relevant
chemical potentials are of order 400-500 $MeV$. Both the strange
quark mass and the $\beta$-equilibrium in conjunction with
electrical neutrality have the effect of producing a mismatch in
the Fermi momenta of the pairing fermions. For increasing mismatch
the BCS pairing mechanism is lost and the system can undergo a
phase transition either to the normal state or to a different
phase.

Let us review how the mismatch in the Fermi momenta is originated.
Consider a massive fermion and a massless one at the same chemical
potential $\mu$. The corresponding Fermi momenta are \be
p_{F_1}=\sqrt{\mu^2-M^2}\approx
\mu-\frac{M^2}{2\mu},~~~~p_{F_2}=\mu\ee Therefore the mismatch is
of order $M^2/\mu$. For the second instance, consider that
$\beta$-equilibrium requires the chemical potential for the
electrons to be the opposite of the chemical potential associated
to the electric charge. Then, the neutrality condition determines
uniquely $\mu_e$ in terms of the other chemical potentials. Since
$\mu_e$ is not arbitrary one is forced to consider the free energy
along the neutrality line rather than long a line of fixed
$\mu_e$. The result of this analysis, at first sight, is a little
bit surprising, since where one has unstable phases moving along
lines of constant $\mu_e$, now one gets stable phases along
neutrality lines \cite{huang:2003ab}.

If we characterize the mismatch in terms of the difference among
effective chemical potentials, call it $2\delta\mu$, the typical
spectrum of quasi-fermions is
$E=\pm\,\delta\mu+\sqrt{(p-\mu)^2+\Delta^2}$. We see that if
$\delta\mu\ge\Delta$ gapless modes are present in correspondence
of momenta $p=\mu\pm\sqrt{\delta\mu^2-\Delta^2}$. The point
$\delta\mu=\Delta$ plays a special role, since the energy cost for
pairing the fermions belonging to Fermi spheres with a mismatch is
$2\delta\mu$ whereas the gain for pairing is $2\Delta$. Therefore
when $\delta\mu>\Delta$ fermions start to lose their BCS pairing.

Examples of the previous situation are the gapless phases g2SC
\cite{huang:2003ab} and gCFL \cite{Alford:2003fq,Alford:2004hz}.
In the g2SC phase the fermion condensate has the same structure as
in the 2SC phase, but there is a mismatch $\delta\mu=\mu_e/2$, and
correspondingly a phase transition from 2SC to g2SC at
$\mu_e=2\Delta$. The analysis shows that the g2SC phase is stable
along the neutrality line. In this particular example the quarks
involved are the quark up and the quark down, so there is no
effect from the strange quark mass, but only from the interplay of
$\beta$-equilibrium and electrical neutrality.

The gCFL has a fermionic condensate given by \be \langle
0|\psi_{aL}^\alpha\psi_{bL}^\beta|0\rangle=\Delta_1\epsilon^{\alpha\beta
1}\epsilon_{ab1}+\Delta_2\epsilon^{\alpha\beta
2}\epsilon_{ab2}+\Delta_3\epsilon^{\alpha\beta 3}\epsilon_{ab3}\ee
whereas in CFL, $\Delta_1=\Delta_2=\Delta_3$. In this case the
mass strange quark mechanism for the mismatch plays the driving
role. In fact, the mismatch among the blue-down and green-strange
quarks is given by $M_s^2/(2\mu)$ and a phase transition between
the CFL and the gCFL (where $\Delta_3>\Delta_2>\Delta_1$) occurs
exactly at $M_s/\mu=2\Delta$. From the analysis made in
\cite{Alford:2004hz} it turns out that gCFL is the favored phase
over 2SC and g2SC up  to about $M_s^2/\mu\approx 130$ $MeV$, for a
choice of the gap $\Delta$ in CFL given by 25 $MeV$.

In presence of a mismatch among the Fermi momenta of the pairing
fermions there is another interesting possibility which has been
considered in the literature, that is the so called LOFF phase
\cite{LOFF} (see also the reviews \cite{reviews}). In this case
the mechanism proposed is a different way of pairing in which each
of the pairing fermions stays close to its own Fermi surface. As a
consequence the pair has a non zero total momentum leading to a
breaking of translational and rotational invariance
\cite{Alford:2000ze}. This phase is particularly interesting since
it may give rise to a crystalline structure. This possibility was
widely explored in \cite{Bowers:2002xr} within the context of a
Ginzburg-Landau expansion. Since the validity of this expansion in
the present context is not completely justified the result of the
analysis was a conjecture about the most favored structure, a face
centered cube. In \cite{Casalbuoni:2004wm} a different
approximation to the problem was proposed. This approximation
holds in a region far from a second order transition, and in this
sense is complementary to the Ginzburg Landau expansion. It was
found that when the BCS phase is lost a LOFF phase with a
crystalline structure corresponding to an octahedron (or a body
centered cube) takes place. This happens at
$\delta\mu=\Delta/\sqrt{2}$. This phase remains the favored one up
to $\delta\mu\approx 0.95\Delta$ where the face-centerd cube
crystal becomes energetically favored. Then this phase persists up
to 1.32$\Delta$, when the system goes back to the normal phase.
According to the authors of ref. \cite{Alford:2004hz} this would
imply (extrapolating the results of ref. \cite{Casalbuoni:2004wm}
obtained in the 2SC case), that the LOFF phase should take place
at about $M_s^2/\mu\approx 120$ $MeV$ up to 225 $MeV$. If this
would be the case the phases 2SC and g2SC would play no role since
increasing $M_s$ one would go from CFL to gCFL through a second
order transition and then via a first order one to the LOFF phase.


\end{document}